\documentstyle[12pt]{article}
\begin{document}
\title{Toward the Collapse of State}
\author{Slobodan Prvanovi\'c and Zvonko Mari\'c \\ Institute of Physics, P.O. Box 57, 11001 Belgrade,\\ Serbia}
\maketitle
\begin{abstract}
The basic concepts of classical mechanics are given in the
operator form. Then, the hybrid systems approach, with the
operator formulation of both quantum and classical sector, is
applied to the case of an ideal nonselective measurement. It is
found that the dynamical equation, consisting of the Schr\"odinger
and Liouville dynamics, produces noncausal evolution when the
initial state of measured system and measuring apparatus is chosen
to be as it is demanded in discussions regarding the problem of
measurement. Nonuniqueness of possible realizations of transition
from pure noncorrelated to mixed correlated state is analyzed in
details. It is concluded that collapse of state is the only
possible way of evolution of physical systems in this case.

\end{abstract}
\section{Introduction}
The correct theory of combined quantum mechanical and classical
mechanical systems has to differ from quantum mechanics (QM) and
classical mechanics (CM) with respect to causality and related
topics. This is because the dynamical equations of QM and CM, taken
separately, cannot lead to such changes of states that can happen in
a (quantum) measurement processes. Quantum and classical mechanics are
causal theories in which pure states can evolve, according to
the appropriate equations of motion, only into pure states, not into
the mixed ones. For a process of nonselective measurement on some
QM system, done by an apparatus which is a CM system, there is a
possibility for transitions from pure to mixed state.

An interesting approach to hybrid systems (consisting of one
quantum and one classical system) was proposed in literature
\cite{1,2}. In short, it uses for states and observables the
direct product of QM and CM representatives. The dynamical
equation there introduced is, let say, superposition of QM and CM
dynamical equations. But, it was objected that this equation of
motion does not save the non-negativity of states, which has to be
unaltered if the theory is supposed to be physically meaningful.
Otherwise, there would be events whose occurrence is characterized
with negative probabilities. However, we shall try to show that
the hybrid systems approach (HSA) is the adequate theoretical
framework for description of an ideal nonselective measurement.

The employed strategy will be the following. Firstly, for a
particular choice of initial state of QM system and measuring
apparatus, which addresses the problem of measurement, it will be
shown that a correlated state, in contrast to the initial, cannot
be pure. Secondly, it will be found that the (dramatic) change of
purity can be formally realized in more than one way; only one of
them will be unphysical for involved negative probabilities. In
order to find what should be taken as the state of this hybrid
system after the beginning of measurement, the subtle analysis is
needed. It should support ones belief that the change of purity is
necessarily followed by the change of this or that property of
state.

We shall keep the argumentation on the physical ground. Precisely,
the necessary requirements to respect the physical meaning
whenever it is possible, and/or to consider only physically
meaningful mathematical entities when physical problems are
discussed, will be sufficient here for finding the other,
physically meaningful possibility for mixed correlated state as
the result. It will become obvious that this state is in
accordance with expected collapse of QM state, as is suggested by
the above (anarchical) title.

Before showing that, we shall propose an operator formulation of
classical mechanics. We shall use it instead of the standard phase
space formulation of CM within the HSA. It will allow us to proceed the
argumentation in more complete way. However, it can be used
separately with some other intentions.

\section{The Operator Description of Classical Mechanics}
The most important features of the well-known phase space formulation
of classical mechanics are: {\bf 1.)}
the algebra of observables is commutative, {\bf 2.)} the equation
of motion is the Liouville equation and it incorporates the Poisson bracket
and {\bf 3.)} pure states
are those with sharp values of position and momentum, the values of
which are, in general, independent. All these will hold for the operator
formulation of CM which we are going to introduce heuristically.

Let the pure states for position, in the Dirac notation, be $\vert
q \rangle $. Similarly, for momentum: $\vert p \rangle $. In
quantum mechanics independence of states is formalized by the use of direct
product. These prescriptions suggest that pure classical states
should be related somehow with $\vert q \rangle \otimes \vert p
\rangle$. Consequently, the operator formulation of classical
mechanics should be looked for within the direct product of two
rigged Hilbert spaces, let say ${\cal H}^q \otimes {\cal H}^p $.
In such a space, one can define an algebra of classical
observables. It is the algebra of polynomials in
$\hat q_{cm} = \hat q \otimes \hat I$ and $\hat
p_{cm} = \hat I \otimes \hat p$ with real coefficients, {\sl etc.}
The elements of this algebra are Hermitian operators and they obviously
commute since $[\hat q_{cm} , \hat p_{cm} ]=0$. Further, one can define
states like in the standard formulation of CM as functions of position and momentum,
which are now operators. Precisely, one can define the pure states as:
$$
\delta (\hat q - q(t))\otimes \delta (\hat p - p(t)) = \int \int
\delta (q-q(t)) \delta (p-p(t)) \vert q \rangle \langle q \vert
\otimes \vert p \rangle \langle p \vert dqdp =
$$
\begin{equation}
=\vert q(t) \rangle \langle q(t) \vert \otimes \vert p(t) \rangle
\langle p(t) \vert .
\end{equation}
The pure and  (noncoherently) mixed states, commonly denoted by
$\rho (\hat q_{cm}, \hat p_{cm}, t)$ in this formulation, are
non-negative and Hermitian operators, normalized to $\delta ^2 (0)$ if
for the same function of real numbers, {\sl i.e.}, for $\rho (q,p,t)$,
it holds that $\rho (q,p,t) \in {\rm \bf R} $,
$\rho (q,p,t) \ge 0 $ and $\int \! \int \rho (q,p,t)\ dq\ dp=1$.
If one calculates the mean values of observables, {\sl e.g.},
$f(\hat q_{cm},\hat p_{cm})$, in state $\rho (\hat q_{cm}, \hat
p_{cm},t))$ by the Ansatz:
\begin{equation}
{{\rm Tr} (f(\hat
q_{cm},\hat p_{cm}) \rho (\hat q_{cm}, \hat p_{cm},t))\over {\rm Tr}
\rho (\hat q_{cm}, \hat p_{cm}, t) },
\end{equation}
then it will be equal to the usually calculated
$\int \int f(q,p) \rho (q,p,t)dqdp$ where $f(q,p)$ and $\rho (q,p,t)$
are the phase space representatives of corresponding observable
and state, respectively. It is easy to see that, due to (1) and
(2), the third characteristic of phase space formulation holds for
the new one as well.

For the criterion of purity we propose the idempotency of state,
up to its norm. This criterion is obviously satisfied for (1) and
it is adequate for the standard formulation of QM. Therefore, we
shall use it for the operator formulation of hybrid systems, too.

The dynamical equation in the new formulation can be defined in accordance
to {\bf 2.)} as:
\begin{equation}
{\partial \rho (\hat q_{cm},\hat p_{cm},t) \over \partial t } =
{\partial H (\hat q_{cm},\hat p_{cm})
\over \partial \hat q_{cm}}
{\partial \rho (\hat q_{cm},\hat p_{cm},t) \over \partial \hat
p_{cm}} -{\partial H (\hat q_{cm},\hat p_{cm}) \over \partial \hat
p_{cm}} {\partial \rho (\hat q_{cm},\hat p_{cm},t) \over \partial
\hat q_{cm}}.
\end{equation}
For the RHS of (3) we shall use the notation $\{ H(\hat q_{cm},\hat
p_{cm}) , \rho (\hat q_{cm},\hat p_{cm},t) \}$.

The standard formulation of classical mechanics appears through the
kernels of the operator formulation in the $\vert q \rangle
\otimes \vert p \rangle$ representation. This, together with (2), can be used as the proof of equivalence of the two formulations. The other important remark is that, after the $\hat q_{cm}$ and $\hat p_{cm}$ have been defined,
each other observable and every state can and have to be expressed as some
function of just these two.

\section{An Outline of the Hybrid Systems App\-roach}

A physical system is called hybrid system if it consists of one
QM and one CM system. Such systems were discussed in [1-7].
Instead of reviewing these articles with purpose of introducing
formalism for hybrid systems, we shall start with the standard
treatment of two QM systems and then, by substituting one quantum
with one classical system, find directly the appropriate
theoretical framework.

The standard formulation of two quantum systems needs the direct product
of two (rigged) Hilbert spaces, let say ${\cal H}_{qm1} \otimes {\cal
H}_{qm2}$. The states of these systems evolve according to the
Schr\"odinger equation with the Hamiltonian $\sum _{\alpha} \hat H
^{\alpha} _{qm1} \otimes \hat H _{qm2} ^{\alpha}$, for which it holds:
$$
{\partial (\sum _{ij} \hat \rho ^{ij}_{qm1}(t) \otimes \hat \rho ^{ij}
_{qm2}(t)) \over \partial t }=
{1\over i\hbar } [ \sum _{\alpha} \hat H ^{\alpha}_{qm1} \otimes
\hat H ^{\alpha}_{qm2},
\sum _{ij} \hat \rho ^{ij}_{qm1}(t) \otimes \hat \rho ^{ij}_{qm2}(t)]=
$$
$$
=\sum _{\alpha ij} {1\over i\hbar } [ \hat H ^{\alpha}_{qm1} ,
\hat \rho ^{ij}_{qm1}(t) ] \otimes {\hat H ^{\alpha}_{qm2} \hat
\rho ^{ij}_{qm2}(t) + \hat \rho ^{ij}_{qm2}(t) \hat H ^{\alpha}_{qm2}
\over 2} +
$$
\begin{equation}
+ \sum _{\alpha ij}{\hat H ^{\alpha}_{qm1} \hat \rho ^{ij}_{qm1}(t) +
\hat \rho ^{ij}_{qm1}(t) \hat H ^{\alpha}_{qm1} \over 2} \otimes
{1\over i\hbar } [ \hat H ^{\alpha}_{qm2} , \hat \rho ^{ij}_{qm2}(t)].
\end{equation}
With $\sum _{ij} \hat \rho ^{ij}_{qm1} (t) \otimes \hat \rho ^{ij}
_{qm2} (t)$ (and more with the one in next expression) we want to
accommodate the notation for states to that type of correlation
which will be discussed below.

Suppose now that the second system is
classical. This means that everything related to this system in
(4) has to be translated into the classical counterparts. Having in
mind the above formulation of CM, we propose:
$$
{\partial (\sum _{ij} \hat \rho ^{ij}_{qm}(t) \otimes \hat \rho ^{ij}
_{cm}(t) ) \over \partial t }=
$$
$$
=\sum _{\alpha ij} {1\over i\hbar } [ \hat H ^{\alpha}_{qm} ,
\hat \rho ^{ij}_{qm}(t)] \otimes {\hat H ^{\alpha}_{cm} \hat \rho
^{ij}_{cm}(t) + \hat \rho ^{ij}_{cm}(t) \hat H ^{\alpha}_{cm} \over 2}+
$$
\begin{equation}
+ \sum _{\alpha ij} {\hat H ^{\alpha}_{qm} \hat \rho ^{ij}_{qm}
(t) + \hat \rho ^{ij}_{qm}(t) \hat H ^{\alpha}_{qm} \over 2 } \otimes
\{ \hat H ^{\alpha}_{cm} , \hat \rho ^{ij}_{cm}(t)\},
\end{equation}
as the dynamical equation. Few explanations follow. The first
system remained quantum mechanical, so its type of evolution is left unaltered. The
Poisson bracket is there instead of the second commutator because classical
systems evolve according to the Liouville equation. It is defined as
in (3); the partial derivatives are with respect to the classical
coordinate and momentum: $\hat q \otimes \hat I$ and
$\hat I \otimes \hat p$. All states and observables,
both QM and CM, appear in the operator form, {\sl i.e.}, hybrid
system is defined in ${\cal H}_{qm} \otimes {\cal H}^q _{cm}  \otimes
{\cal H}^p _{cm}$. ({\sl Nota bene}, the coordinate and momentum of quantum  and classical systems are operators acting in ${\cal H}_{qm}$ and ${\cal H}^q _{cm}  \otimes {\cal H}^p _{cm}$, respectively.)
Some justifications of (5) we shall give in due
course.

Similar equations, in the c-number formulation of CM, were
proposed in [1-4]. There one can find the whole variety of
requests that has to be imposed on the equation of motion for hybrid
systems which will not be reviewed here. We just mention that
the equation proposed in [1-3] is antisymmetric, while the one in
\cite{4} is not.

More discussions of the same subject one can find in
\cite{5,6}. The starting point there was that the formalism of hybrid
systems should have all mathematical properties of QM and CM (see
\cite{6} for details) and it was concluded that such formalism cannot
exist. Rather than as a critique, we understand this result as an
indication that the HSA is on a right track. Namely, we do not expect
from the appropriate formalism to posses all mathematical properties
being the same as in quantum and classical mechanics. On the contrary, we expect that the correct theory of hybrid systems will differ from these two
mechanics with respect to the causality of evolution and, consequently,
all other related topics. More precisely, in some cases the hybrid
systems equation of motion should lead to the noncausal evolution.
The example we have in mind, as we have mentioned, is a process of
(quantum) measurement.

It was objected in \cite{2,3,7} that the HSA dynamical equation
does not save the non-negativity of states. Our intention is to
show, with a subtle analysis of process of measurement, that this
need not to be so, {\sl i.e.}, the non-negativity of states can be
saved. This comes from our belief that after finding some
dynamical equation as the source of noncausal evolution, what will
be the case for (5), one should accept any kind of instruction, of
course, if there is some, since, on the first place, one would be
faced with the problem in which way it should be solved. That is,
this type of dynamical equations, we believe, should be approached
in different, more careful manner than it is usually the case
because it is not so straitforward job to solve them. On the other
hand, it will be enough to apply some arguments, that are of the
same kind as are those which  qualify non-negative states as
meaningless, and to find acceptable states. This will become clear
latter. At this place, let us just mention that the noncausal
evolution of CM system alone occurred in a treatment of classical
mechanics by the inverse Weyl transform of the Wigner function;
see \cite{8} for details.

\section{The Process of Measurement}

Usually, it is said that the measuring apparatus is classical system. The
formalism of hybrid systems becomes then the natural choice for the 
representation of process of (quantum) measurement. We
shall consider the nonselective measurement within the operator
formulation of HSA by taking that the states of measured QM system and
measuring apparatus evolve under the action of $H_{qm} (\hat q \otimes \hat I \otimes \hat I , \hat p \otimes \hat I \otimes \hat I ) + H_{cm} (\hat I \otimes \hat q \otimes \hat I , \hat I \otimes \hat I \otimes \hat p ) + V_{qm} (\hat q \otimes \hat I \otimes \hat I , \hat p \otimes \hat I \otimes \hat I )\cdot V_{cm} (\hat I \otimes \hat q \otimes \hat I , \hat I \otimes \hat I \otimes \hat p )$. To simplify the expressions, we shall use $ \hat H_{qm} \otimes \hat I _{cm} + \hat  I _{qm} \otimes \hat H_{cm} + \hat V _{qm}
\otimes \hat V _{cm}$ as the notation for this Hamiltonian. The measured observable is
$\hat V_{qm} = \sum _i v_i \vert \psi _i \rangle \langle \psi _i
\vert \otimes \hat I \otimes \hat I $. It is necessary that $[\hat
H_{qm} , \hat V_{qm} ]=0$ because, if the quantum system before
the measurement was in one of the eigenstates of the measured
observable, say $\vert \psi _i \rangle$, then this system would
not change its state during the measurement. Then, $\hat H_{qm}$ can
be diagonalized in the same basis: $\hat H_{qm} = \sum _i h_i
\vert \psi _i \rangle \langle \psi _i \vert \otimes \hat I \otimes
\hat I $. For the CM parts of Hamiltonian it is reasonable to
assume that they do not cause periodic motion of the pointer. We
shall not specify the Hamiltonian in more details because we are
interested only in discussions related to the form of state after
the beginning of measurement.

For the initial state of quantum system we shall take the pure state
$\vert \Psi (t_o) \rangle$ and for the pointer of apparatus we shall
take that initially it is in the state with sharp values of position
and momentum, let say $q_o$ and $p_o$, so the state of hybrid
system at the moment when measurement starts is $\hat \rho _{qm} (t_o)
\otimes \hat \rho _{cm} (t_o) = \vert \Psi (t_o) \rangle \langle \Psi
(t_o) \vert \otimes \vert q_o \rangle \langle q_o \vert \otimes \vert
p_o \rangle \langle p_o \vert $. Of course, the problem of measurement
demands $\vert \Psi (t_o) \rangle$ to be superposition $\sum _i c_i
(t_o) \vert \psi _i \rangle$.

Due to the interaction term in Hamiltonian, the state of composite system will become correlated - the CM parts of state will depend somehow on the eigenvalues of $\hat V _{qm}$. Let us use the notation $\sum _{ij} \hat \rho
^{ij}_{qm} (t) \otimes \hat \rho ^{ij} _{cm}(t)$ in order to allow
the analysis of, {\sl a priori}, possible situation in which the CM
parts of state can depend on two different eigenvalues of $\hat V _{qm}$. With this notation, and the above for Hamiltonian, the dynamics of measurement becomes represented with:
$$
{\partial (\sum _{ij} \hat \rho ^{ij}_{qm} (t) \otimes \hat \rho
^{ij} _{cm}(t)) \over \partial t }=
$$
$$
=\sum _{ij}  {1\over i\hbar } [\hat H_{qm} , \hat \rho ^{ij}_{qm}
(t)] \otimes \hat \rho _{cm}^{ij} (t) + \sum _{ij} {1\over i\hbar
}[\hat V_{qm} , \hat \rho ^{ij}_{qm} (t)] \otimes \hat V_{cm} \hat
\rho _{cm}^{ij} (t) +
$$
\begin{equation}
+\sum _{ij} \hat \rho ^{ij}_{qm} (t) \otimes \{ \hat H _{cm},
\hat \rho _{cm}^{ij} (t) \} + \sum _{ij} {1\over2} (\hat V _{qm} \hat
\rho ^{ij}_{qm} (t) + \hat \rho ^{ij}_{qm} (t) \hat V _{qm}) \otimes
\{ \hat V _{cm}, \hat \rho _{cm}^{ij} (t) \},
\end{equation}
where $\hat H _{cm} , \hat V _{cm}$ and $ \rho _{cm}^{ij} (t)$ are derived in the Poisson bracket with respect to $\hat q \otimes \hat I$ and $\hat I \otimes \hat p$ that act in ${\cal H}^q _{cm} \otimes {\cal H}^p _{cm}$.

The solution of this dynamical equation will represent the state of 
hybrid system at $t>t_o$ and the search for it can start by
noticing that the CM terms $\hat \rho _{cm}^{ii} (t)$, attached to
the quantum mechanical terms with equal indices $\hat \rho
_{qm}^{ii} (t)$ (which we shall call diagonal terms), are $\hat
\rho _{cm}^{ii} (t) = \vert q_i (t) \rangle \langle q_i (t) \vert
\otimes \vert p_i(t) \rangle \langle p_i(t) \vert $, where the
indices in $\vert q_i (t) \rangle $ and $\vert p_i (t) \rangle $ underline 
 dependence on one eigenvalue of $\hat V_{qm}$. Being guided by
this dependence of each CM bra and ket of $\hat \rho _{cm}^{ii}
(t)$ on one eigenvalue of $\hat V _{qm}$, as the candidate for
correlated state we shall consider the coherent mixture:
\begin{equation}
\sum _{ij} c_{ij} (t) \vert \psi _i \rangle \langle \psi _j \vert
\otimes \vert q_i (t) \rangle \langle q_j (t) \vert \otimes \vert
p_i(t) \rangle \langle p_j(t) \vert .
\end{equation}
There are two other candidates for correlated state. The first is:
\begin{equation}
\sum _{ij} c_{ij} (t) \vert \psi _i \rangle
\langle \psi _j \vert \otimes \vert q_{ij} (t) \rangle \langle q_{ij}
(t) \vert \otimes \vert p_{ij} (t) \rangle \langle p_{ij} (t) \vert ,
\end{equation}
where the indices in $\vert q_{ij} (t) \rangle$ and $\vert p_{ij}
(t) \rangle $ stand to represent dependence on two eigenvalues of
$\hat V_{qm}$ in the form ${1\over 2} (v_i +v_j)$. The same holds
for $\langle q_{ij} (t) \vert$ and $\langle p_{ij} (t) \vert$. The motivation
for this comes from the symmetrization of QM sector in front of the Poisson bracket on the RHS of (5). The terms $\hat \rho _{cm}^{ij} (t)$ of
(8) are diagonal with respect to the eigenbasis of $\hat q_{cm}$ and $\hat
p_{cm}$ for each pair of indices, while these terms of (7) for
$i\ne j$ are not. As the third candidate for correlated state we
shall consider the noncoherent mixture:
\begin{equation}
\sum _i \vert c_i (t_o) \vert ^2
\vert \psi _i \rangle \langle \psi _i \vert \otimes \vert q_i (t)
\rangle \langle q_i (t) \vert \otimes \vert p_i (t) \rangle \langle
p_i (t) \vert .
\end{equation}
All three states have the same diagonal terms $\hat \rho _{qm}^{ii} (t) \otimes \hat \rho _{cm}^{ii} (t)$. The difference
between these states is in the CM $i\ne j$ terms.  Each ket and bra
of $ \hat \rho _{cm}^{ij} (t)$, $i\ne j$, in (7) depends on only one
eigenvalue of $\hat V_{qm}$, in (8) they depend on two eigenvalues
and in expression (9) there are no such terms.

The state (7) is designed to represent as pure, non-negative and
Hermitian correlated state as is the initial state and it has
nondiagonal QM terms (with respect to the basis $\vert \psi _i
\rangle$) as the state $\vert \Psi (t_o) \rangle \langle \Psi
(t_o) \vert$. (The state is taken to be pure if it is idempotent up to the norm: $\hat \rho ^2 = \delta ^2 (0) \cdot \hat \rho $.) The purity of (7) rests on the same type of time development (dependence on one $v_i$) of
$\vert q_i (t) \rangle$ and $\vert p_i (t) \rangle$, no matter do
they belong to $\hat \rho _{cm}^{ij} (t)$ with $i=j$ or with $i\ne j$. But, the following holds. The initial state of the apparatus is diagonal with respect to the eigenbasis of $\hat q_{cm}$ and $\hat p_{cm}$. To ``create'' the nondiagonal terms from it in the form which ensures purity, one would need  to introduce operators that do not commute with $\hat q_{cm}$ and
$\hat p_{cm}$ to act on CM states. One would need to take some
other dynamical equation instead of (5) as well. That dynamical
equation should use commutator for both subsystems, like it is the
case for (4). If one would do that, then, in a treatment of the apparatus,  one would neglect the requirements {\bf 1.)} and {\bf 2.)} which are the part of definition of classical systems (see Sec. 2). This type of reasoning would be {\sl a la} von Neumann's approach to measurement process where the apparatus and measured system are both treated as quantum systems. Instead of going in that direction, we are considering here the apparatus as 
classical system, defined in the above given way. By this we avoid
the well known problems that arise with states such is (7).
(According to (7) there could be a superposition of pointers state
which is unobserved. Then, the problem of measurement, as we
understand it, is to explain why and describe how the state
similar to (7) collapses to the state similar to (9).)

The less descriptive and more formal way to look for a solution is
to assume that the time dependence of evolved state is as
represented by (7). Then, by substituting (7) in (6) in order to
verify this, we find a contradiction. Namely, the CM $i\ne j$
terms of (7) do not commute with $\hat q_{cm}$ and $\hat p_{cm}$
for $t\ne t_o$, so then they are not functions of only these
observables. The partial derivatives
${\partial \over \partial \hat q_{cm} }$ and ${\partial \over
\partial \hat p_{cm} }$ from the Poisson bracket ``annihilate'' the CM
nondiagonal elements of (7) for $t>t_o$ when they act on them. For
instance:
\begin{equation}
{\partial \over \partial \hat q } \vert q_i (t) \rangle \langle q_j
(t) \vert = {\partial \over \partial \hat q } \delta (\hat q - q_i
(t)) \cdot \delta _{i,j},
\end{equation}
($t>t_o$) and similarly for $\vert p_i (t) \rangle \langle p_j (t)
\vert$ under the action of ${\partial \over \partial \hat p_{cm} }$.
Thus, for the CM $i\ne j$ terms of (7) the RHS of (6) vanishes for $t>t_o$,
while the LHS is not equal to zero by assumption.

Let us stop for a moment and put few remarks. An immediate
consequence of the fact that (7) does not satisfy (6) is that the
initial purity of state is lost due to the established
correlation. This is confirmed by considerations of (8) and (9).
These two states do satisfy (6), but they are both mixed - they
are not idempotent up to the norm:
$\hat \rho ^2 \ne \delta ^2 (0) \cdot \hat \rho $. This property
is plausible for (9). For (8) it is enough to notice that in $\hat
\rho ^2$ there is,  for example, term $\vert \psi _i \rangle \langle
\psi _i \vert \otimes \vert q_{ij} (t) \rangle \langle q_{ij} (t)
\vert \otimes \vert p_{ij} (t) \rangle \langle p_{ij} (t) \vert$
which is not present in $\hat \rho$. Therefore, the hybrid systems
dynamical equation produces in this particular case a 
noncausal evolution: pure noncorrelated state transforms in
some mixed correlated state (which is to be found).
This is the crucial difference between
(5) and the Schr\"odinger and Liouville dynamics that appear within
it.

One can convince oneself, by looking at (8) and (9), that purity
is not the only property of initial state that changes
instantaneously at the moment when interaction begins. Obviously,
there are no $i\ne j$ terms in (9) the meaning of which is that
the QM part of (9), in difference to the initial one, is diagonal
with respect to the basis
$\vert \psi _i \rangle$. On the other hand, the state (8) is not
non-negative operator for all $t>t_o$, while the initial state is.
For all states that are not non-negative operators one can
construct properties - events, that would be ``found'' with
negative probabilities if they would be measured. In order to
construct such a property for (8), it is helpful to notice that
the CM parts of $i\ne j$ terms of (8) are regular states of CM
systems, they are different from those with $i=j$ and they are
accompanied by the QM ``states'' with vanishing trace. (By regular
we mean {\sl per se} realizable since they are diagonal and
``states'' stands here, and would be better to stay in all similar
cases, because they can only be interpreted as impossible.)

For the related negative probabilities, states which are not
non-negative operators should be qualified as meaningless and,
since they appeared in the HSA, there were objections on its
relevance for physics. In what follows, we want to show that these
probabilities are not unavoidable here. In other words, our
intention is to rehabilitate the HSA and this will manifest itself
in finding formal support for physically meaningful state (9),
that it should be taken as the solution, not the unphysical state
(8). The arguments have to be in accordance with physics since the
experience makes one to be unsatisfied with (8) and, of course,
the HSA is aimed to formalize behaviour of physical systems. The
first argumentation, being based on the validity of (10), will
continue the analysis of (7). The second discussion, concentrated
on (8) and unrelated to (10), will again designate that (9) is the
proper solution, but, in difference to the first one, it will be
proceeded in more interpretational than formal manner.

Our insistence on (7) rests on the fact that one can look on it as
on a trial state. It is the perfect choice for a trial state
because it has the same physically relevant characteristics as the
initial state and it is equal to the initial state for $t =t_o$,
{\sl i.e.}, for $t \rightarrow t_o$ (7) approaches the initial
state without any change when these characteristics are
considered. Moreover, the need for a trial state comes from the
absence (up to our knowledge) of some rule that would prescribe
how to manage the change of idempotency. After being substituted
on the RHS of dynamical equation, trial state will indicate the
appropriate type of time transformation. Then, by minimal
modifications of this state, intended to adapt it to that type,
desired correlated state will be found.

The RHS of (6) for the CM $i\ne j$ terms of (7) vanishes for all $t>t_o$ according to (10). Exclusively for $t=t_o$ the CM $i\ne j$
terms of (7) can be expressed as functions of only $\hat q_{cm}$
and $\hat p_{cm}$ since $q_i (t_o ) =q_o$ and $p_i (t_o )=p_o $
for all $i$. Only for this moment the RHS of (6) for CM $i\ne j$
terms of (7) does not vanish. Therefore, one concludes that the CM
parts of $i\ne j$ terms has to be constant after the instantaneous
change at $t_o$, {\sl i.e.}, instead with those of (7), the QM
nondiagonal terms have to be coupled with the time independent CM
terms for all $t>t_o$. This is how dynamical equation
designates that (8) should not be taken as the solution. What one
has to do, if one wants to accommodate (7) to deduced time
independence of the CM $i\ne j$ terms, is to take for these terms (for
$t>t_o$) some operators that do not involve time. Then, in order
to satisfy (6), that operators should not be expressible as some
functions of (only available) $\hat q _{cm}$ and $\hat p _{cm}$. On the other
hand, with these operators one should not change neither the
Hermitian character nor the non-negativity of state since nothing
asks that. The resulting state, of course, has to be impure
because any change of the CM $i\ne j$ terms of (7) affects its
idempotency. In this way, (9) will be obtained as the appropriate
solution.

Having in mind the functions of $\hat q_{cm}$, $\hat p_{cm}$ and
operators that do not commute with these two, one may want not to
accept (10). For the sake of mathematical rigor, let us clear up
this. The CM nondiagonal terms of (7) cannot be expressed as some
functions depending only on $\hat q_{cm}$ and $\hat p_{cm}$, but
they can be expressed as some functions of these two if, firstly,
the number of the operators available is increased and, secondly,
there is some non-commutativity among them. How this functions
would look like depends on these new operators. Since there are neither motivations nor instructions for their introduction coming from physics, 
they can be introduced liberately. More precisely, these operators do not
represent anything meaningful and they need not to enclose any
known mathematical structure. For instance, $\vert q_i (t) \rangle
\langle q_j (t) \vert$ can be expressed as $\exp ( {1\over a} (q_i
(t) - q_j (t) ) \hat \pi ) \delta (\hat q - q_j (t))$, where $\hat
\pi $ is not to be confused with the CM momentum since it acts in
${\cal H}^q$, not in ${\cal H}^p$, and $\langle q \vert \hat \pi
\vert q' \rangle = a {\partial \delta (q-q') \over {\partial q }}
$. Here, $a$ can be anything, it need not to be equal to $-i\hbar$
as in quantum mechanics. The other (even more pathological) example is the
following. Since the CM nondiagonal dyads do not commute with
$\hat q_{cm}$ and $\hat p_{cm}$, they can be used as the new
operators, {\sl e.g.}, $\vert q_i (t) \rangle \langle q_j (t)
\vert = F(q_i (t))^{-1} F(\hat q ) \vert q_i (t) \rangle \langle
q_j (t) \vert$. This shows that these nondiagonal dyads can be
expressed as functions depending on $\hat q _{cm}$, $\hat p _{cm}$
and uncountably many other arguments - all nondiagonal dyads,
where $F$ can be any function. With these two examples we wanted
to justify the need to bound considerations of CM in operator form
to functions of only $\hat q_{cm}$ and $\hat p_{cm}$. On the other
hand, the request to discuss purity of state of the hybrid system
has risen the need to consider nondiagonality (with respect to the basis  $\vert q \rangle \otimes \vert p \rangle$) of CM state. When
these two meet in dynamical equation, with expressions like (10)
nothing unusual was done: the derivation of an entity, which is
not some function of that with respect to which it is derived, has
zero as a result. If one says that the LHS of (10) is just defined
by the RHS of (10), then it should be noticed that (10) does not
contradict any of the calculational rules of CM and QM because in
the standard formulation of classical mechanics there is no
possibility for realization of nondiagonality, while in the
standard formulation of quantum mechanics there is no necessity
for restriction to commutativity. Anyhow, let us proceed by
supposing that one is not willing to accept (10) and/or that one
finds the given support for (9) as not enough convincing.

Even without (10), one is not free of contradiction if (7) is
taken to be the solution. Due to the symmetrization of QM sector,
on the RHS of (6), in front of the second Poisson bracket, there
are two eigenvalues of $\hat V_{qm}$ coming from $\hat \rho
_{qm}^{ij} (t)$ ($i\ne j$) of (7). Because of this, the assumption
that each ket and bra of $\hat \rho _{cm}^{ij} (t)$ ($i\ne j$) of
(7) depends on only one eigenvalue of $\hat V_{qm}$ is
contradicted. As it seems, to introduce non-commuting operators in
${\cal H}^q _{cm} \otimes {\cal H}^p _{cm}$, and/or to slightly
modify (6), would not be enough to avoid some contradiction
connected to (7) when it is seen as the result of evolution.
However, it is not our intention to go in these directions because
it would be against the purpose of this article.

After discarding (7), one concludes that each ket and bra of
$\hat \rho _{cm}^{ij} (t)$ ($i\ne j$) would depend on two eigenvalues
of $\hat V_{qm}$ coming from $\hat \rho _{qm}^{ij} (t)$ ($i\ne j$) for
 $t>t_o$ if there would be $\hat \rho _{qm}^{ij} (t)$ ($i\ne j$) for
that times at all. Therefore, the most important step in solving dynamical
equation for the above Hamiltonian is to find what happens with the
initial QM state at the moment when interaction begins. Then it will be
almost trivial problem to find the state of hybrid system at latter times. Or,  more precisely, in the presence of $\vert \psi _i \rangle \langle
\psi _j \vert$ ($i\ne j$) for $t>t_o$ is the origin of dilemma: (8)
or (9), the meaning of which is that by the assumed linearity of evolution,
in a case when it is noncausal, one excludes the physical meaning of evolved
state, and {\sl vice versa}.

From this point, our strategy for defense of the HSA from
objections that it might be unphysical is in showing that one
finds it unphysical only after one has previously decided to
prefer formal, rather than physical arguments and, moreover, only
after one has neglected statements (being, by the way, of the same
sort as those used for disqualification) that lead to physically
meaningful state. Let us be more concrete. To find (8) it was
necessary to start with more formal assumption that the
nondiagonality of QM part of state, with respect to the eigenbasis
of $\hat H_{qm}$ and $\hat V_{qm}$, has not changed at the moment
when purity of state has changed. Opposite to this is to assume
that the diagonality of QM part of state, with respect to the basis
which is privileged at that time, has not changed. Before the
moment $t_o$, the QM part of state has been diagonal with respect
to the eigenbasis of that observable for which $\vert \Psi (t_o )
\rangle$ is the eigenstate. Only this basis can be characterized
as privileged for that time because the corresponding observable
has been used for preparation. For physics, each other basis, including
the eigenbasis of $\hat H_{qm}$ and $\hat V_{qm}$, is less
important, {\sl i.e.}, their significance comes from mathematics,
not from physics - they can be used just to express the same state in
different manners. After the moment $t_o$ privileged basis is the
eigenbasis of $\hat V_{qm}$ (and $\hat H_{qm}$) because this 
observable is measured. So, instead of claiming that the 
nondiagonality with respect to the basis which is going to become
privileged should not change, one can claim that the diagonality with
respect to the actually privileged basis should not change. These
statements express two different types of reasoning: the first one
concentrated on the formal aspect of the operators representing
states (leading to (8)), while the other one  concerned about the
meaning (leading to (9)).

If the mentioned nondiagonality of QM part of initial state has survived
$t_o$, then, according to (6), there would
be the CM systems in (realizable) states $\hat \rho ^{ij}_{cm}
(t)$ coupled to the QM nondiagonal terms, as is given by (8). But,
the probability of event $\hat I \otimes \vert q_{ij} (t) \rangle
\langle q_{ij} (t)\vert \otimes \vert p_{ij} (t) \rangle \langle
p_{ij} (t) \vert $ for the state (8) is equal to zero for all
$t>t_o$, where $i \ne j$. Neither apparatus would be in any of the
states $\hat \rho ^{ij}_{cm} (t)$ with $i \ne j$ after the
beginning of measurement. (This is not the case for $i=j$.) So, if
the statements about probability are of any importance, before
proclaiming (6) as unadequate for it does not save the
non-negativity of initial state, one should accept that in the
states $\hat \rho ^{ij}_{cm} (t)$ ($i\ne j$) neither apparatus
would be. The consequence of this is that the assumption of
survived QM nondiagonal terms is not correct. In physics, where
the probability is a significant concept, just found is enough to
conclude that (9) should be taken as solution. Simultaneously by
finding that (8) is unphysical, one finds why it is so: it is
unphysical because some states of CM systems that are not exhibited
by any apparatus are kept in the representation of state of hybrid
system. By taking this into account, {\sl i.e.}, by reexpressing
(8) with this in mind, one will find (9) as the proper state of
hybrid system.

Finally, the validity of the hybrid systems dynamical equation can
be verified on situations for which it is easy to say what behavior
is desired. For example, the hybrid systems dynamical equation
gives the standard one-to-one evolutions of QM and CM subsystems
when the interaction term in Hamiltonian is absent. In this case 
evolved states are of the same purity and non-negativity as 
initial states. Moreover, for the above given Hamiltonian and the
initial state of hybrid system $\sum _i \vert c_i (t_o) \vert ^2
\vert \psi _i \rangle \langle \psi _i \vert \otimes \vert q_o
\rangle \langle q_o \vert  \otimes \vert p_o \rangle \langle p_o
\vert $, evolved state is not unphysical, it is (9). These
examples justify the hybrid systems dynamical equation as the
proper one. So, it is likely that this holds for the case addressing 
the problem of measurement.

\section{Concluding remarks}

Without an operator formulation of classical mechanics, the
analysis of the problem of measurement in the hybrid systems
approach would not be complete. Firstly, this formulation enabled
us to consider pure correlated state and then, after finding that
such state cannot satisfy dynamical equation, to conclude that
this dynamical equation produces noncausal evolution: when pure
initial state of quantum system is not an eigenstate of the
measured observable, initial state of hybrid system, which is also
pure, necessarily and instantaneously transforms in some mixed
correlated state. Secondly, when it was not so obvious how
dynamical equation should be solved, the operator formulation
offered support for one particular way.

The choice of a state representing hybrid system after the 
beginning of measurement is important since appropriateness of the
HSA for physics depends on it. Both states that do satisfy
dynamical equation for the given Hamiltonian are same regarding
the impurity and absence of CM $i\ne j$ terms, so the essential
part of physical meaning is one and the same. Only the way of
expressing these differs from (8) to (9). For their properties,
perhaps it would not be wrong to say that (9) is the physical
result of hybrid systems dynamics and that (8) is a physically
unacceptable mathematical solution.

The third usefulness of the operator formulation of classical
mechanics is in that it allows one to design, let say, a dynamical
model of instantaneous decoherence. Namely, in the resulting
proposal of HSA, the partial derivations in the Poisson bracket
change the CM nondiagonal terms at $t_o$ (if the initial state is seen
as (7) with $t=t_o$) and then obstruct their further time development according to (10), {\sl i.e.}, these derivations annihilate CM nondiagonal
terms. So, in this proposal, the dynamics is the cause of
collapse. The reduction of quantum mechanical state is
the consequence of disappearance of classical mechanical $i\ne j$
terms. The part of interpretation of (8), which is meaningful from
the point of view of everyday experience, has lead to the same
conclusion: terms $\hat \rho ^{ij} _{qm} (t)$ vanish because to them related  and {\sl per se} realizable events $\hat \rho ^{ij} _{cm} (t)$ cannot occur. In another words, the reason for decoherence of QM state in case of a
measurement lies in the Liouville equation. It is linear only in
probability densities within the framework of commutative
operators that represent position and momentum of classical
systems, in difference to the Schr\"odinger equation which is
linear in both: the probability densities and the probability
amplitudes.

In almost the same manner as the action of projectors has
described the measurement in standard quantum mechanics, the
action of partial derivations do it here. If one compares the
standard formulation of QM and the operator formulation of HSA,
one finds them similar for they treat decoherence as instantaneous
process. They differ since decoherence is dynamical here. The
operator formulation of HSA in this way answers one question
aroused in quantum mechanics: how the collapse should be
described. But, there is another, more important question: why it
happens. The hybrid system approach does not ask for some {\sl ad
hoc} concepts to explain the collapse of state; the non-negativity
of probabilities is enough. Because of the non-negativity of
probabilities, the collapse of state is the only possible way of
evolution for physical systems in the considered case and it is as
ordinary as the one-to-one evolutions are in other cases. If one
wants to stay within the formulation of QM in one Hilbert space,
then the HSA puts the projection postulate on more solid ground.
It is not related to the consciousness of the observer, but to the
non-negativity of probabilities.

The non-negativity of probabilities is, and should be,
incorporated among the first principles of any physical theory.
The hybrid system approach differs from classical and quantum
mechanics only in that this principle should be invoked not just
at the beginning, when the initial state is chosen, but for the
moments at which states lose purity as well. This rule offers
substitution of our search for a solution and it is not in
contradiction with these two mechanics. There are no such moments
when only Schr\"odinger or Liouville equation is solved within the
Hilbert space and phase space, respectively, so there is no rule
which would be contradicted. If it is represented (like some kind
of superselection rule) in ${\cal H}_{qm} \otimes {\cal H}^q _{cm}
\otimes {\cal H}^p _{cm}$ as a restriction to consider only states
that are non-negative operators, then there would be only two
possibilities for a correlated state in the analyzed case: the
coherent mixture (7) and the noncoherent mixture (9). The state
(9) would follow immediately after finding that (7) cannot satisfy
the equation of motion. (There is strong similarity between this and the way  of solving the Maxwell quations where only physicaly meaningful solution is retained.)

Roughly speaking, the procedure of solving differential equations
consists in two steps. The first one is to find all functions that
satisfy it (if there is any) and the second is, if there are more
than one function, to select one by imposing some condition. The
most often used is the Cauchy condition. Adapted to the present
framework, it reads: the state at later times is the one which for
$t=t_o$ becomes equal to the initial state.  With this condition
one wants to express assumed continuity of state. The state (8)
obviously follows in this way and, since this state is unphysical,
the HSA shows that the state of physical systems in considered
case has to evolve discontinuously.  From our point of view, this
strongly recommends the HSA for a theory of combined classical and
quantum systems.

The objections addressing the relevance of HSA for physics are
closely related to the application of the Cauchy condition in, let
say, careless manner. We believe that it is not correct to
take it as the unique supplementary condition and that it is not appropriate  to impose it without noticing that something dramatic happens with the
initial state at the moment when evolution begins. If one
would disregard the unavoidable change of purity of initial state
treating it as unimportant, then one would go out of physics from
the very beginning. Moreover, then one cannot discuss the physical
meaning of solution at the end because it would make such
consideration inconsistent. Only after finding that (according to
the discussion based on (7) and (10)) the initial state has
changed instantaneously and discontinuously, one should apply the
Cauchy condition for then it is adequate because the further evolution is causal and in all aspects continuous. If this, the rule to invoke the
non-negativity of probability for the moments at which states lose
purity and (10) are new at all, these rules are the slightest
possible modifications of the previously used ones. Or, perhaps,
they are just the accommodation of standard rules to new
situations.

Needless to say, the state (9) is in agreement with what is usually
expected to happen when the problem of measurement is considered in an
abstract and ideal form. To each state of the measured quantum system,
which are the eigenstates of the measured observable, corresponds one pointer
position and momentum. The $i$-th eigenvalue of measured observable
occurs with probability $\vert c_i (t_o) \vert ^2$ and, as was said, (9) takes
place immediately after the apparatus in state $\vert q_o \rangle
\otimes \vert p_o \rangle $ has started to measure $\hat V _{qm}$ on
the system in pure state $\vert \Psi (t_o) \rangle$, which can be
seen as $\sum _i c_i (t_o) \vert \psi _i \rangle $.

Once noticed, the departure from strict causality would
also be noticed in (all) other aspects as some strange feature. For
example, in \cite{6} it was found that, so called, universal privileged
times in dynamics of hybrid systems appear. Here, $t_o$ is such a moment. In
contrast to opinion expressed there, we believe that this is a rather
nice property of the approach. Namely, for the described process, and
all other that can be treated in the same way, pure state can evolve
into noncoherent mixture, while noncoherent mixture cannot evolve
into coherent mixtures - pure states, {\sl i.e.}, when the non-negativity of probability is respected, such processes are irreversible. This means that for  them the entropy can only increase or stay constant. Then, the distinguished moments of the increase of entropy can be used for defining an arrow of time.

\end{document}